\documentclass{article}
\usepackage{graphicx}
\usepackage[english]{babel}
\usepackage{listings} 
\newcommand{\beq}{\begin{equation}}
\newcommand{\eeq}{\end{equation}}
\newcommand{\bea}{\begin{eqnarray}}
\newcommand{\eea}{\end{eqnarray}}
\newcommand{\real}{{\sf I}\kern-.12em{\sf R}}
\newcommand{\comp}{{\sf I}\kern-.50em{\sf C}}
\newcommand{\unity}{{\sf I}\kern-.54em{\sf 1}}

\newcommand{\stringa}{\ttfamily\lstinline}
\def\cod#1{{\stringa!#1!}}

\title{
\includegraphics[width=0.35\textwidth]{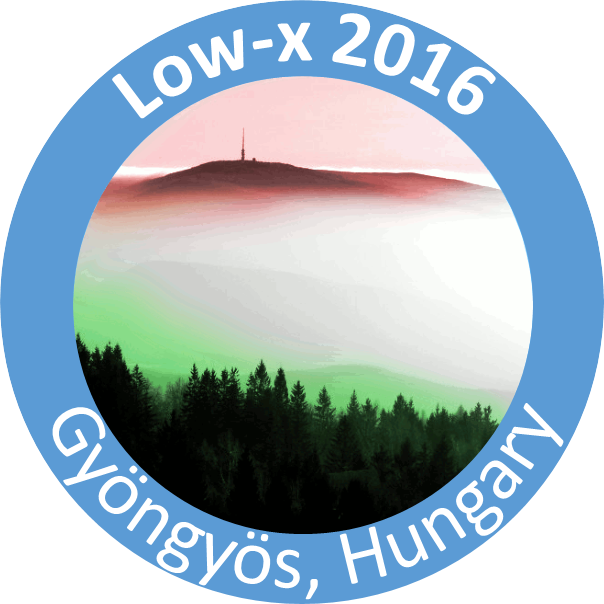}
\\[1cm]
High energy effects in multi-jet production at LHC}
\author{{F.~Caporale$^{1,2}$, 
         F.G.~Celiberto$^{1,2,3,4 \, 
          \footnote{Speaker.}}$,
         G.~Chachamis$^{1,2}$,} 
         \\ 
         D.~Gordo~G{\'o}mez$^{1,2}$,
        {B.~Murdaca$^4$, A.~Sabio~Vera$^{1,2}$}\\[1ex]
$^1$Instituto de F{\'i}sica Te{\'o}rica UAM/CSIC,\\ 
     Nicol{\'a}s Cabrera 15, Madrid, Spain\\
$^2$Universidad Aut{\'o}noma de Madrid, 28049 Madrid, Spain\\
$^3$Dipartimento di Fisica, Università della Calabria,\\ 
     Arcavacata di Rende, 87036 Cosenza, Italy\\
$^4$Istituto Nazionale di Fisica Nucleare, 
     Gruppo Collegato di Cosenza, \\ 
     Arcavacata di Rende, 87036 Cosenza, Italy
}

\begin{document}

\fontfamily{lmss}\selectfont
\maketitle

\begin{abstract}
{We study differential cross sections 
          for the production of three and four jets in 
          multi-Regge kinematics, 
          the main interest lying on azimuthal angle
          dependences. The theoretical setup 
          is the jet production from a single BFKL
          ladder with a convolution of two/three 
          BFKL Green functions, where two forward/backward
          jets are always tagged in the final state. 
          Furthermore, we require the tagging of one/two further
          jets in more central regions of the detectors 
          with a relative separation in rapidity. 
          We found, as result, that the dependence on 
          transverse momenta and rapidities 
          of the central jets can be considered as a
          distinct signal of the onset of BFKL dynamics.
          }
\end{abstract}

\section{Introduction}

The study of semi-hard processes 
in the high-energy (Regge) limit
represents an ultimate research field for perturbative QCD, 
the Large Hadron Collider (LHC) 
providing with an abundance of data. 
Multi-Regge kinematics (MRK),
which prescribes final state objects 
strong ordered in rapidity,
is the key point for the study 
of multi-jet production at LHC energies.
In this kinematical regime, 
the Balitsky-Fadin-Kuraev-Lipatov 
(BFKL) approach, at leading (LL) 
\cite{Lipatov:1985uk,Balitsky:1978ic,
Kuraev:1977fs,Kuraev:1976ge,Lipatov:1976zz,Fadin:1975cb} 
and next-to-leading 
(NLL)~\cite{Fadin:1998py,Ciafaloni:1998gs} accuracy, 
is the most powerful tool
to perform the resummation of large logarithms 
in the colliding energy 
to all orders of the perturbative expansion.
This formalism was successfully applied 
to lepton-hadron Deep Inelastic Scattering at HERA 
(see, {\it e.g.}~\cite{Hentschinski:2012kr,
Hentschinski:2013id}) in order to study quite inclusive 
processes which are not that suitable though 
to discriminate between BFKL dynamics 
and other resummations. 
The high energies reachable at the LHC, however, 
allow us to investigate processes 
with much more exclusive 
final states which could, in principle, 
be only described by the BFKL framework, making it possible 
to disentangle the applicability region of the approach. 
So far, Mueller--Navelet jet production~\cite{Mueller:1986ey} 
has been the most studied process.
Interesting observables associated to this reaction are 
the azimuthal correlation momenta which, however, 
are strongly affected by collinear contaminations.
Therefore, new observables independent 
from the conformal contribution
were proposed in~\cite{Vera:2006un,Vera:2007kn} 
and calculated at NLL 
in~\cite{Ducloue:2013bva,Caporale:2014gpa,Caporale:2015uva,
Celiberto:2015yba,Celiberto:2016ygs,
Ciesielski:2014dfa,Angioni:2011wj,Chachamis:2015crx},  
showing a very good agreement 
with experimental data at the LHC. 
Nevertheless, Mueller-Navelet configurations 
are still too inclusive to perform MRK precision studies. 
Pursuing the goal to further and deeply probe the BFKL dynamics 
by studying azimuthal decorrelations where the transverse momenta
of extra particles introduces a new dependence, 
we proposed new observables for semi-hard processes 
which can be thought as a generalization 
of Mueller-Navelet jets\footnote{Another interesting 
and novel possibility, the detection of two charged light hadrons:
$\pi^{\pm}$, $K^{\pm}$, $p$, $\bar p$ 
having high transverse momenta and 
separated by a large interval of rapidity,
together with an undetected soft-gluon radiaton emission, 
was suggested in~\cite{Ivanov:2012iv} 
and studied in~\cite{Celiberto:2016hae}.}. 
These processes are inclusive 
three-jet~\cite{Caporale:2015vya,Caporale:2016soq}
and four-jet production~\cite{Caporale:2015int,Caporale:2016xku}.

\section{Multi-jet production}

\begin{figure}[h]
\centering
\includegraphics[scale=0.295]{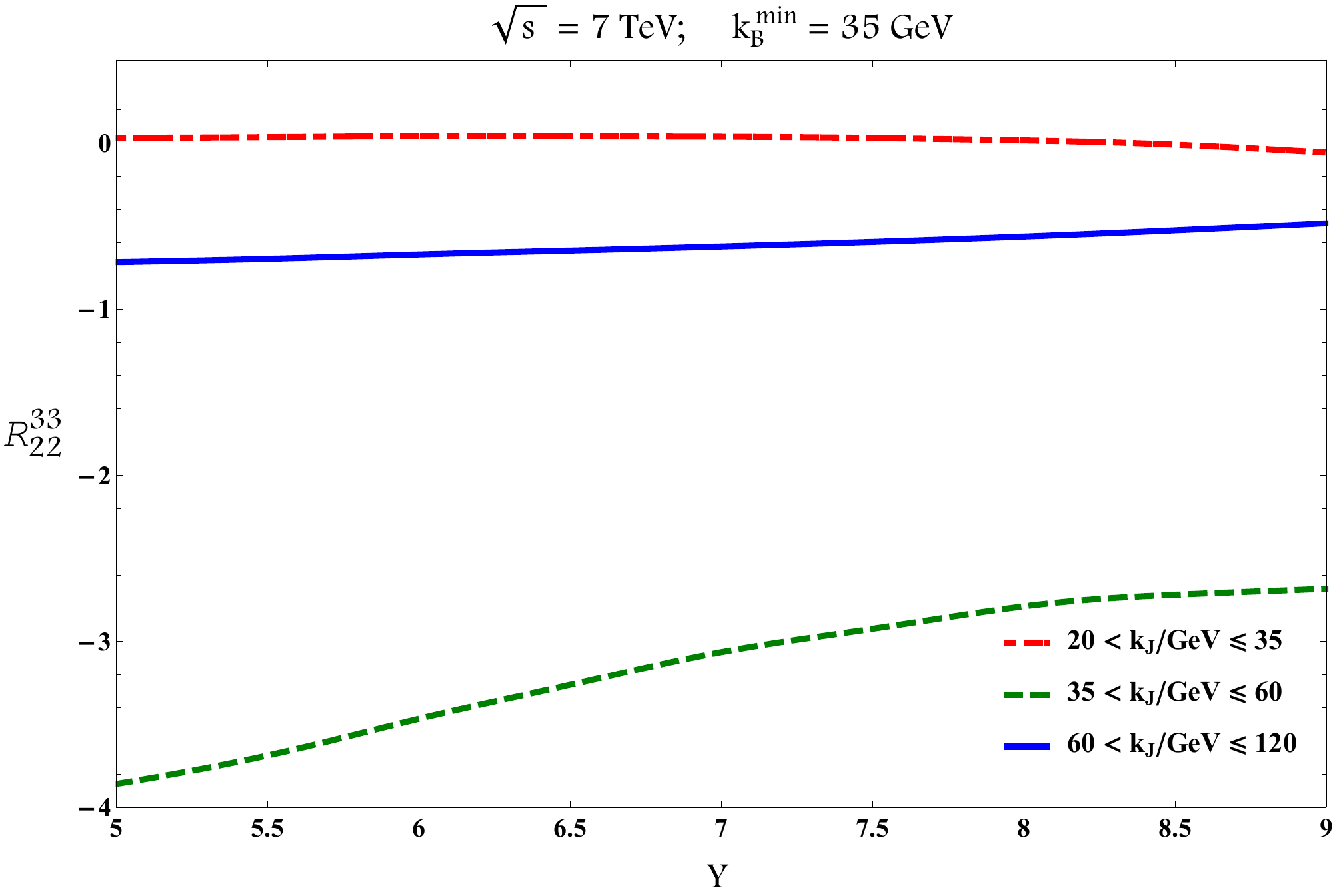}
\includegraphics[scale=0.295]{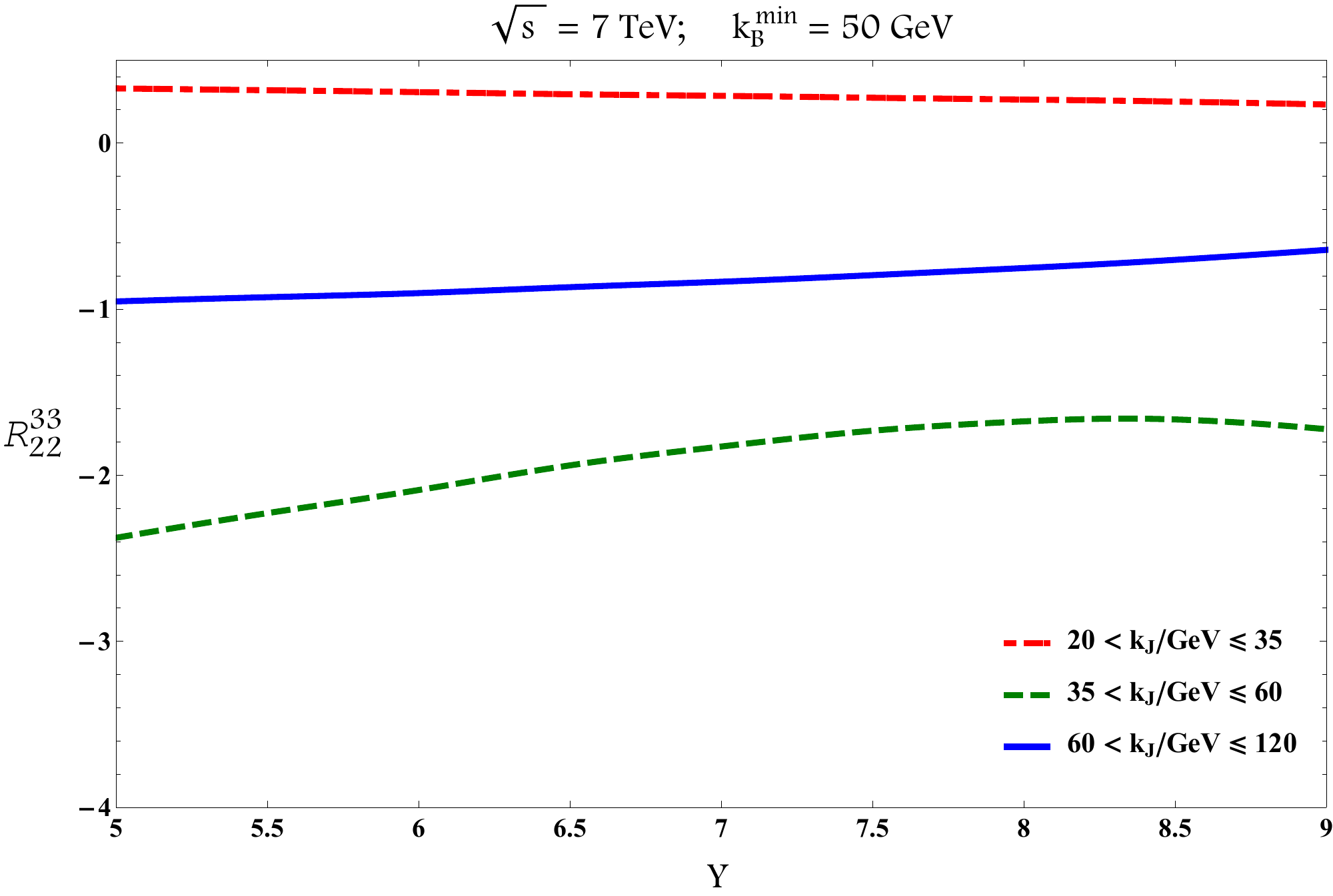}

\includegraphics[scale=0.295]{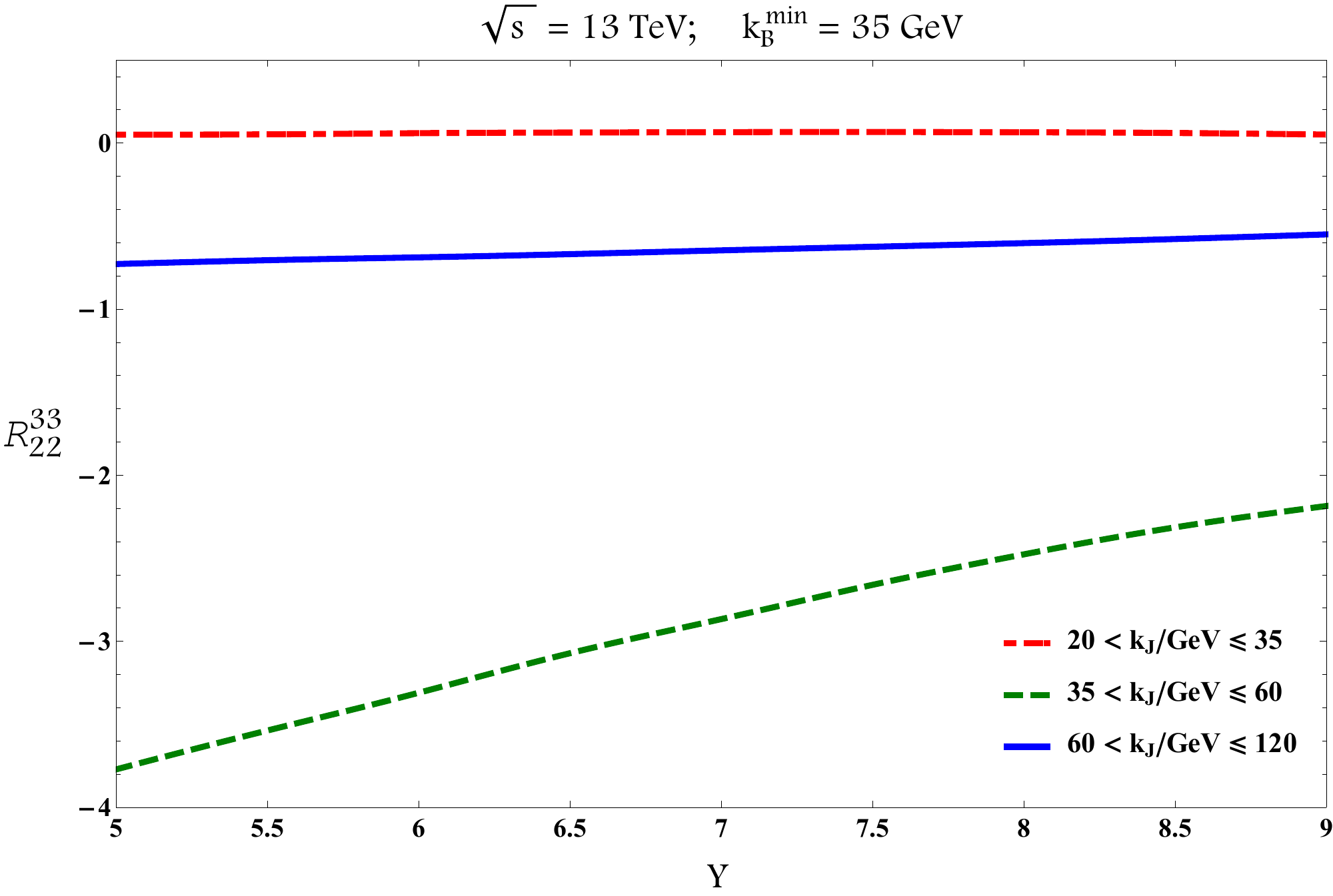}
\includegraphics[scale=0.295]{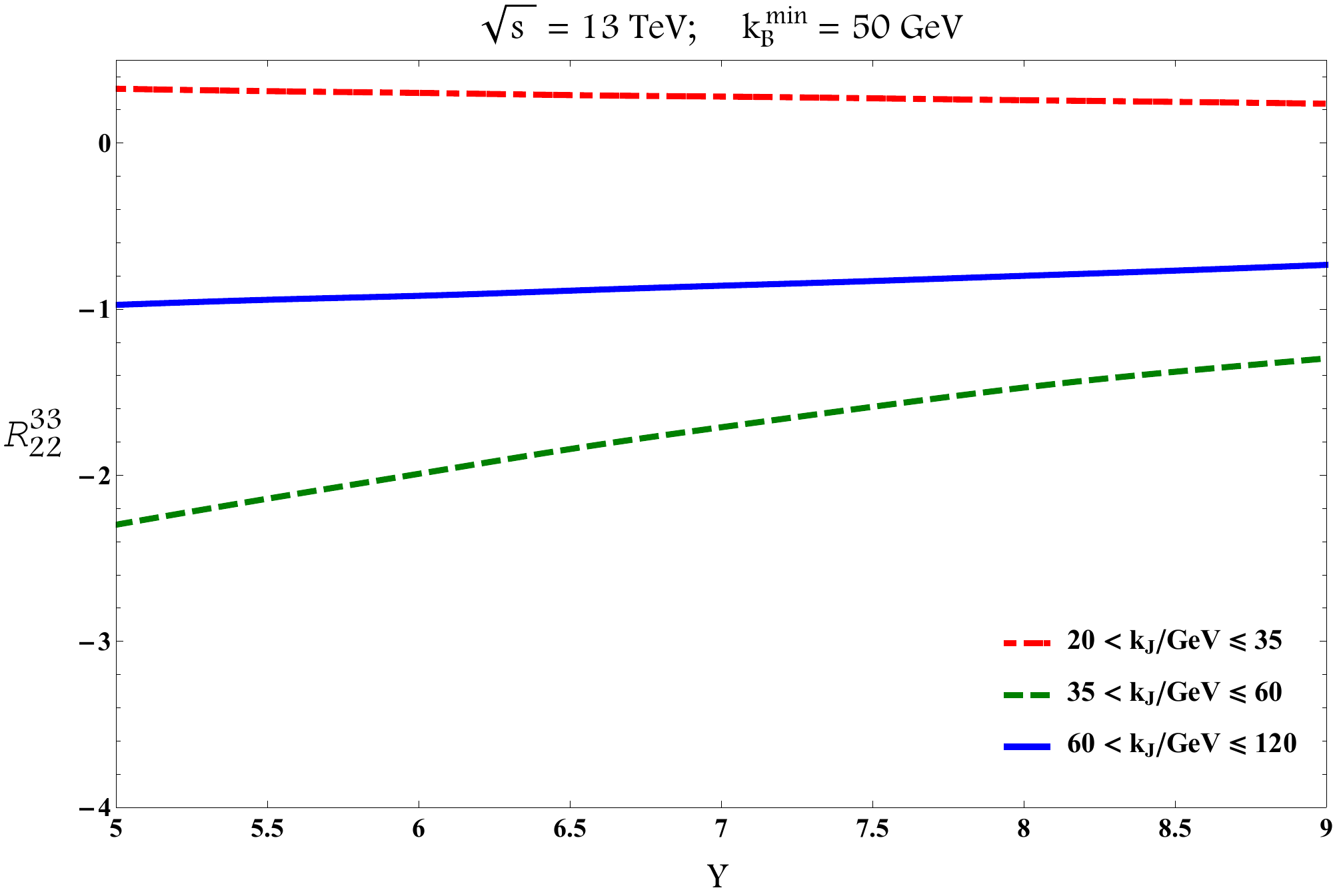}
\caption{$Y$-dependence of $R^{33}_{12}$ for $\sqrt{s} = 7, 13$ TeV 
and $k_{B,\rm min}$ = 35 GeV (left column)
and $k_{B,\rm min}$ = 50 GeV (right column). 
$k_{A,\rm min}$ is equal to 35 GeV, 
while the rapidity of the central jet 
is fixed to $y_J = (y_A + y_B)/2$.}
\label{fig:3jet}
\end{figure}
\begin{figure}[h]
\centering
\includegraphics[scale=0.169]{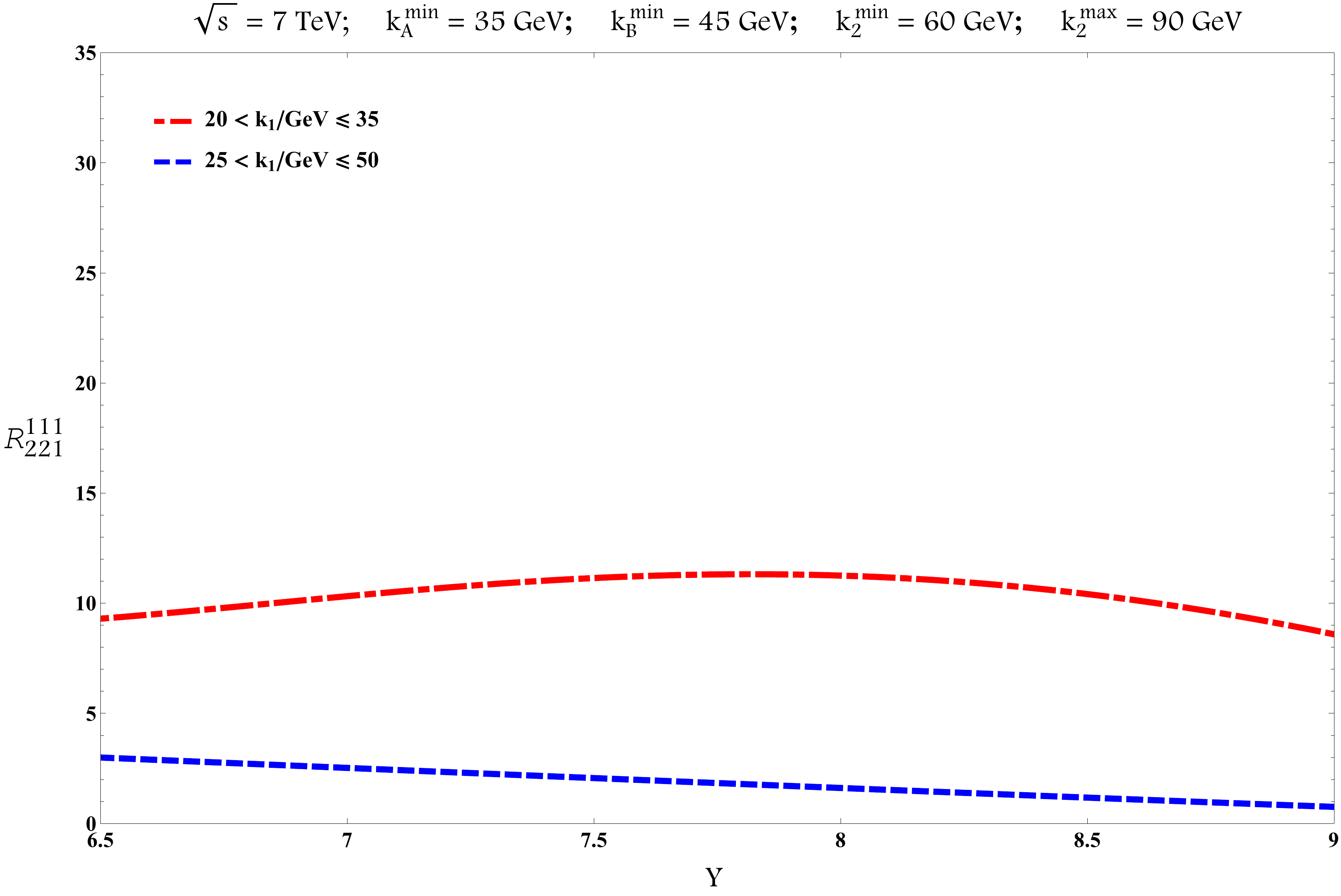}
\includegraphics[scale=0.169]{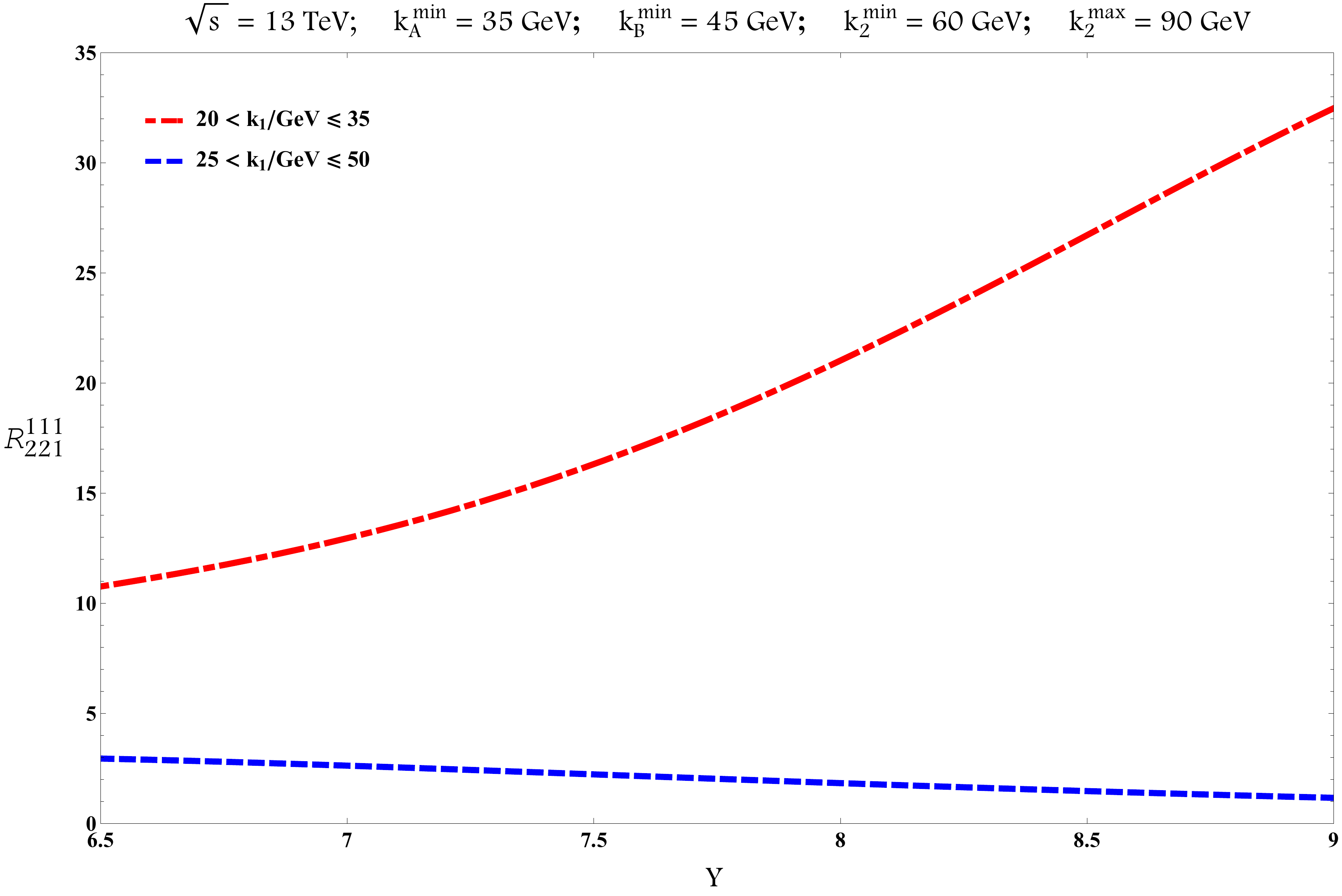}

\includegraphics[scale=0.169]{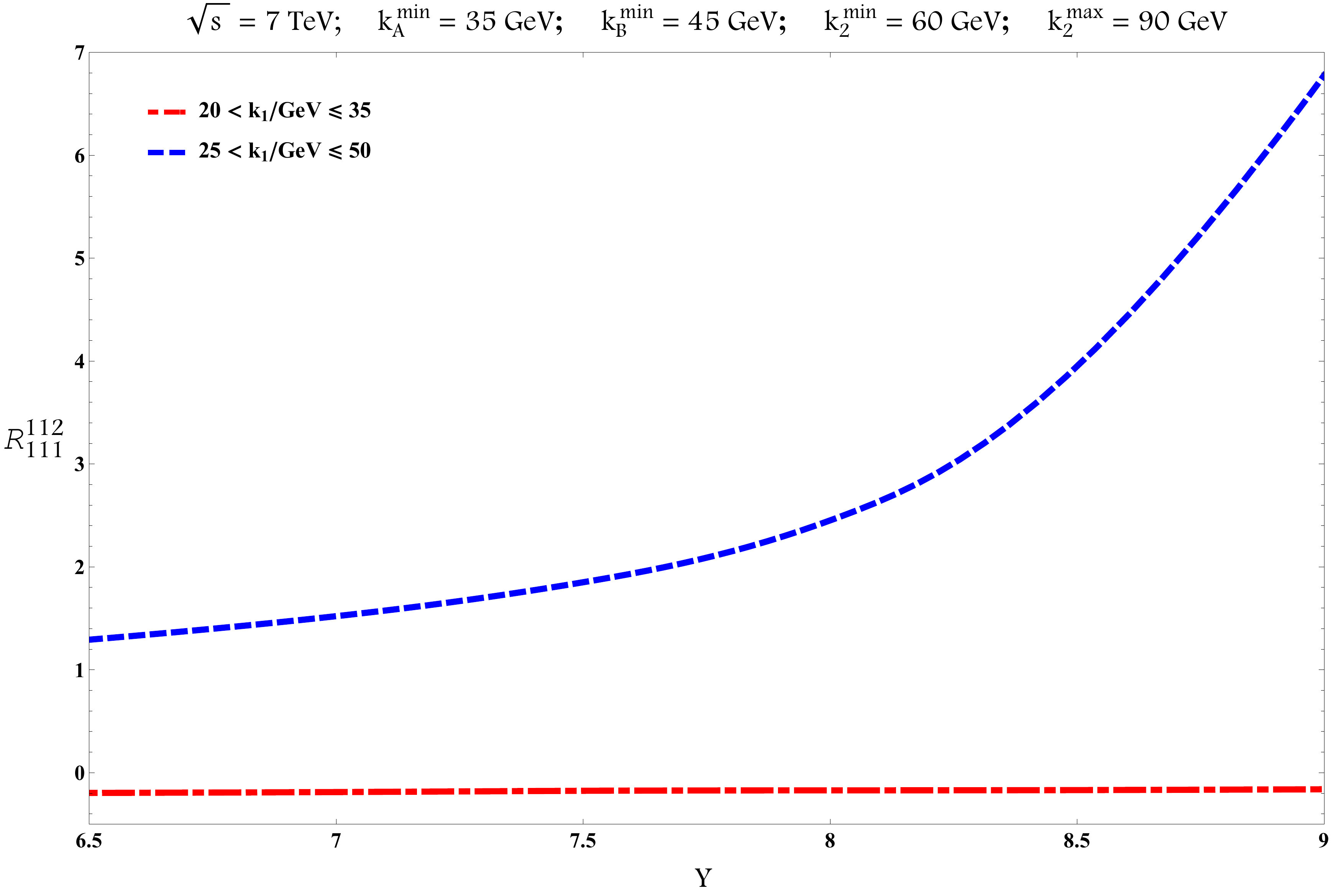}
\includegraphics[scale=0.169]{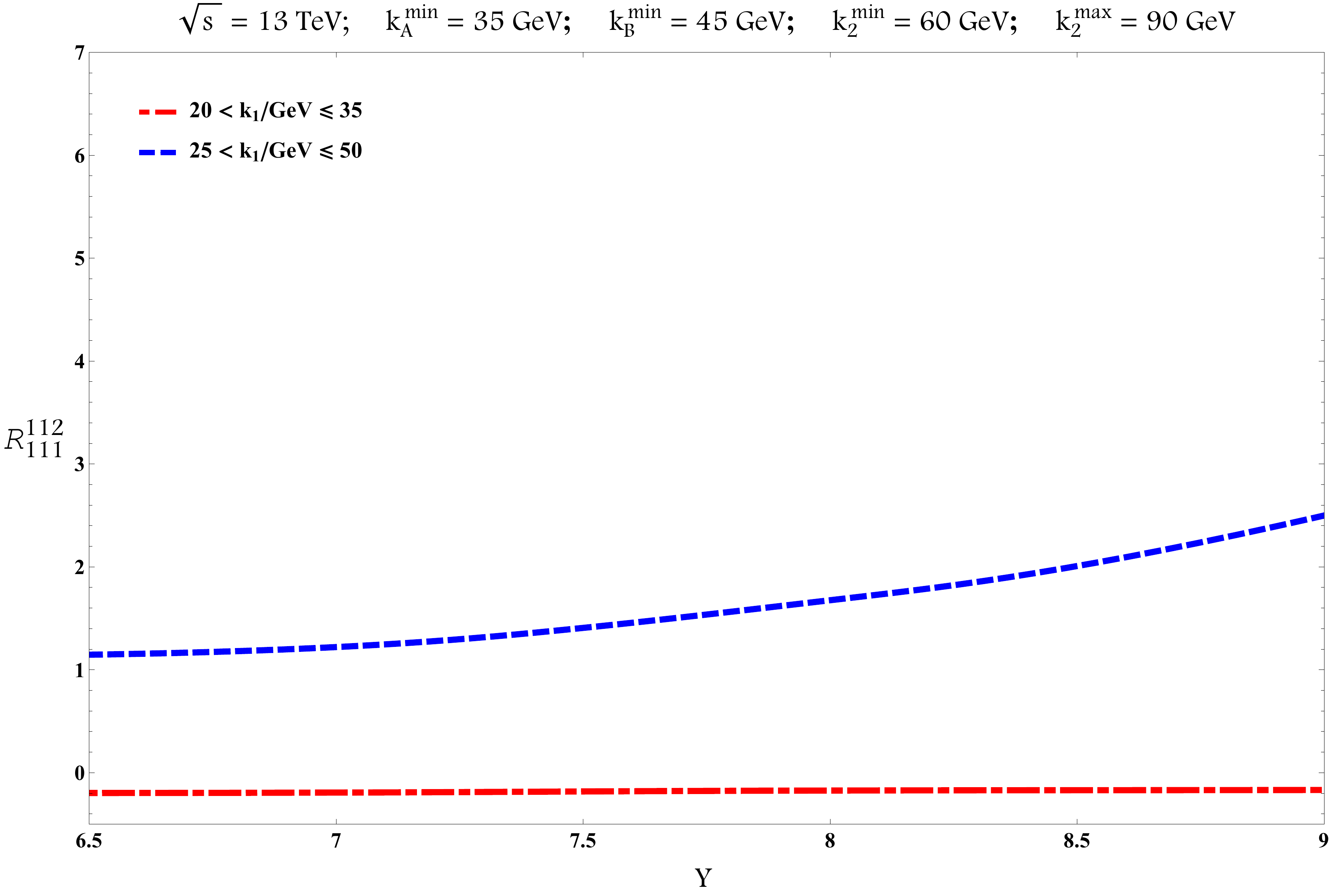}
\caption{$Y$-dependence of $R^{111}_{221}$ and $R^{112}_{111}$
for $\sqrt{s} = 7$ TeV (left column)
and for $\sqrt{s} = 13$ TeV (right column). 
The rapidity interval between a jet and the closest one 
is fixed to $Y/3$.}
\label{fig:4jet}
\end{figure} 

The class of processes under exam 
is the inclusive hadroproduction of $n$ jets in the final state, 
well separated in rapidity so that 
$y_i > y_{i+1}$ according to MRK, 
and with their transverse momenta $\{k_i\}$ 
lying above the experimental resolution scale,
together with an undetected gluon radiaton emission.
With the aim to generalize the azimuthal ratios $R_{nm}$ 
defined in the Mueller--Navelet jet configuration, 
we propose new, generalized azimuthal observables 
by taking the projection of the differential cross section 
$d\sigma^{n-{\rm jet}}$
on all angles, so having the general expression 
given in Eq.~(3) of~\cite{Celiberto:2016vhn}:
\begin{equation}
\mathcal{C}_{M_1 \cdots M_{n-1}} =
\left\langle 
 \prod_{i=1}^{n-1} \cos\left(M_i \, \phi_{i,i+1}\right)
\right\rangle = \hspace{-0.1cm} 
\int_0^{2\pi} \hspace{-0.4cm} d\theta_1 
\cdots \hspace{-0.1cm} 
\int_0^{2\pi} \hspace{-0.4cm} d\theta_n
\prod_{i=1}^{n-1} \cos\left(M_i \, \phi_{i,i+1}\right)
d\sigma^{n-{\rm jet}}
\end{equation}
where $\phi_{i,i+1} = \theta_i - \theta_{i+1} - \pi$, 
and $\theta_i$ is the azimuthal angle of the jet $i$.
From a phenomenological perspective, 
we want to provide predictions compatible 
with the current and future experimental data.
To this purpose, we introduce the kinematical cuts 
already in place at the LHC by integrating 
$\mathcal{C}_{M_1 \cdots M_{n-1}}$ 
over the momenta of all tagged jets in the form
\begin{equation}\label{Cm_int}
C_{M_1 \cdots M_{n-1}} =
\int_{y_{1,\rm min}}^{y_{1,\rm max}}\hspace{-0.25cm}dy_1
\int_{y_{n,\rm min}}^{y_{n,\rm max}}\hspace{-0.25cm}dy_n
\int_{k_{1,\rm min}}^{\infty}\hspace{-0.25cm}dk_1
\cdots
\int_{k_{n,\rm min}}^{\infty}\hspace{-0.25cm}dk_n
\delta\left(y_1-y_n-Y\right)
{\cal C}_n 
\end{equation}
where the most forward and the most backward jet rapidities 
are taken in the range delimited by 
$y_1^{\rm min} = y_n^{\rm min} = -4.7$  and 
$y_1^{\rm max} = y_n^{\rm max} = 4.7$, keeping their difference 
$Y = y_1 - y_n$ fixed.
From a theoretical point of view, 
it is important to improve the stability of our 
predictions (see~\cite{Caporale:2013uva} for a related discussion).
This can be done by removing the zeroth conformal spin contribution 
responsible for any collinear.
For this reason, we introduce the ratios 
\begin{equation}
R^{M_1 \cdots M_{n-1}}_{N_1 \cdots N_{n-1}} \equiv 
\frac{C_{M_1 \cdots M_{n-1}}}{C_{N_1 \cdots N_{n-1}}}
\end{equation}
where $\{M_i\}$ and $\{N_i\}$ are positive integers.

We performed the numerical computation of the ratios 
$\mathcal{R}^{MN}_{PQ}$ both 
in \textsc{Fortran} and in \textsc{Mathematica} 
(mainly for cross-checks).
The NLO MSTW 2008 PDF sets~\cite{MSTW:2009} were used 
and for the strong coupling $\alpha_s$ we chose 
a two-loop running coupling setup 
with $\alpha_s\left(M_Z\right)=0.11707$. 
We made extensive use of the integration routine 
\cod{Vegas}~\cite{VegasLepage:1978} 
as implemented in the \cod{Cuba} 
library~\cite{Cuba:2005,ConcCuba:2015}.
Furthermore, we used the \cod{Quadpack} 
library~\cite{Quadpack:book:1983}
and a slightly modified version 
of the \cod{Psi}~\cite{RpsiCody:1973} routine.

In Fig.~\ref{fig:3jet} we show the dependence on $Y$ 
of the $R^{33}_{22}$ ratio, 
characteristic of the 3-jet process, 
for $\sqrt{s} = 7$ and $13$ TeV, for two different kinematical cuts 
on the transverse momenta $k_{A,B}$ of the external jets 
and for three different ranges of the 
transverse momentum $k_J$ of the central jet.

In Fig.~\ref{fig:4jet} we show the dependence on $Y$ 
of $R^{111}_{221}$ and $R^{112}_{111}$ ratios, 
characteristic of the 4-jet process, 
for $\sqrt{s} = 7$ and $13$ TeV, for asymmetrical cuts 
on the transverse momenta $k_{A,B}$ of the external jets
and for two different configurations 
of the central jet transverse momenta $k_{1,2}$. 

A comparison with predictions for these observables from
fixed order analyses as well as from the BFKL inspired 
Monte Carlo {\bf\cod{BFKLex}}~\cite{Chachamis:2011rw,
Chachamis:2011nz,Chachamis:2012fk,
Chachamis:2012qw,Caporale:2013bva,
Chachamis:2015zzp,Chachamis:2015ico} 
is underway.

\section{Conclusions \& Outlook}

We studied ratios of correlation functions of products 
of azimuthal angle difference cosines in order to study 
three- and four-jet production at hadron colliders. 
The dependence on the transverse momenta and rapidities 
of the central jet(s) represent a clear signal of the BFKL dynamics. 
For future works, more accurate analyses are needed: 
higher order effects and study of
different configurations for the rapidity range 
of the two central jets, together with the analysis of the effect 
of using different PDF parametrizations. 
It would be also interesting to calculate
our observables using other approaches 
not based on the BFKL approach and to
test how they differ from our predictions. 
The comparison with experimental data will help 
to disentangle the region of applicability of the BFKL approach, 
therefore we strongly encourage experimental collaborations 
to study these observables in the next LHC analyses.

\vspace{0.25cm}
\begin{flushleft}
{\bf \large Acknowledgements}
\end{flushleft}
GC acknowledges support from the MICINN, Spain, 
under contract FPA2013-44773-P. DGG acknowledges 
financial support from `la Caixa'-Severo Ochoa doctoral fellowship.
ASV and DGG acknowledge support from the Spanish Government 
(MICINN (FPA2015-65480-P)) and, together with FC and FGC, 
to the Spanish MINECO Centro de Excelencia Severo Ochoa Programme (SEV-2012-0249). 
FGC thanks the Instituto de F{\'\i}sica Te{\'o}rica 
(IFT UAM-CSIC) in Madrid for warm hospitality.

\end{document}